\documentclass[proof]{pasj00}
\draft

\begin{document}
\SetRunningHead{A. Seko et al.}{Gas-to-Dust Ratio in Star-Forming Galaxies at $z \sim 1.4$}

\title{Constraint on the Gas-to-Dust Ratio in Massive Star-Forming Galaxies at $z \sim 1.4$}

\author{
Akifumi \textsc{Seko},\altaffilmark{1}
Kouji \textsc{Ohta},\altaffilmark{1}
Bunyo \textsc{Hatsukade},\altaffilmark{2}
Kiyoto \textsc{Yabe},\altaffilmark{2}
Tomoe \textsc{Takeuchi},\altaffilmark{1}
and
Daisuke \textsc{Iono} \altaffilmark{2,3}
}
\altaffiltext{1}{Department of Astronomy, Kyoto University, Kitashirakawa-Oiwake-Cho, Sakyo-ku, Kyoto 606-8502}
\email{seko@kusastro.kyoto-u.ac.jp}
\altaffiltext{2}{National Astronomical Observatory of Japan, 2-21-1 Osawa, Mitaka, Tokyo 181-8588}
\altaffiltext{3}{The Graduate University for Advanced Studies (SOKENDAI), 2-21-1 Osawa, Mitaka, Tokyo 181-0015}



\maketitle

\begin{abstract}
We carried out $^{12}$CO($J=2-1$) observations toward three star-forming galaxies on the main sequence at $z\sim1.4$ with the Nobeyama 45m radio telescope. 
These galaxies are detected with {\it Spitzer}/MIPS in 24 $\mu\mathrm{m}$, {\it Herschel}/SPIRE in 250 $\mu\mathrm{m}$ and 350 $\mu\mathrm{m}$, and their gas metallicity, derived from optical emission line ratios based on near infrared spectroscopic observations, is close to the solar metallicity. 
Although weak signal-like features of CO were seen, we could not detect significant CO emission. The dust mass and the upper limits on the molecular gas mass are $(3.4-6.7) \times 10^{8}\ M_\odot$ and $(9.7-14) \times 10^{10}\ (\alpha_\mathrm{CO}/4.36)\ M_\odot$, respectively. The upper limits on the gas-to-dust ratios at $z\sim1.4$ are $150-410$ which are comparable to the gas-to-dust ratios in local galaxies with similar gas metallicity. A line stacking analysis enables us to detect a significant CO emission and to derive an average molecular gas mass of $1.3\times10^{11}\ M_\odot$ and gas-to-dust ratio of 250. This gas-to-dust ratio is also near that in local galaxies with solar metallicity. These results suggest that the gas-to-dust ratio in star-forming galaxies with solar metallicity does not evolve significantly up to $z\sim1.4$. 
By comparing to a theoretical calculation, a rapid increase of the dust mass in an earlier epoch of galaxy evolution is suggested.
\end{abstract}

\section{Introduction}
\label{sec: intro}
Revealing the properties of the interstellar matter in high redshift galaxies is crucial for the understanding of galaxy formation and evolution. Since stars form from gas, the gas mass and its mass fraction in high redshift galaxies are key parameters to trace galaxy evolution. Dust plays an important role in the formation of hydrogen molecules and in the cooling of interstellar medium, and its mass also reflects the star formation history in a galaxy (e.g., \cite{key-D3}). It is particularly important to unveil the nature of the interstellar medium in galaxies at $z=1\sim2$, because this coincides with the peak of the cosmological evolution of star formation rate density in the universe (e.g., \cite{key-H2}). 
Since a few years, high-sensitivity radio telescopes enable us to detect the CO emission from the so-called "main sequence" of star-forming galaxies at $z = 1-2.5$ (e.g., \cite{key-D2}, \yearcite{key-D1}; \cite{key-T1}, \yearcite{key-T2}; \cite{key-G1}, \yearcite{key-G3}). The molecular gas mass fractions ($f_\mathrm{mol} = M_\mathrm{mol}/(M_\mathrm{mol} + M_\ast)$) in star-forming galaxies at $z = 1-2.5$ turn out to be $\sim30-50$ \% and are significantly higher than those in the present-day massive spiral galaxies.
Furthermore, new data from {\it Herschel} allows us to investigate the dust emission from main sequence galaxies up to $z \sim 2$ (e.g., \cite{key-E1}; \cite{key-M2}). 

\authorcite{key-M1} (\yearcite{key-M1}, \yearcite{key-M2}) estimate the gas masses in star-forming galaxies at $z\sim2$ from the dust mass with the assumption that the galaxies at this redshift follow the same relation between gas-to-dust ratio and metallicity as local galaxies; the gas-to-dust ratios are larger in galaxies with lower gas metallicity (e.g., \cite{key-L1}; \cite{key-S1}; \cite{key-R2}).  
\citet{key-M2} find that star-forming galaxies at $z\sim2$ show higher molecular gas mass fraction than nearby galaxies, and that these fractions are comparable to those derived from CO observations by \citet{key-D1}. 
Moreover, \citet{key-M2} find a dependence of molecular gas mass fraction on the stellar mass of a galaxy; the molecular gas mass fraction decreases with increasing stellar mass, which also agrees with the result from the CO observations at $z\sim1-1.5$ (\cite{key-T2}).

Theoretical calculations, however, predict a redshift evolution of gas-to-dust ratio (e.g., \cite{key-I2}; \cite{key-C1}). \citet{key-I2} calculated the evolution of the masses of gas, metal, and dust in a galaxy considering star formation, dust core growth, metal accretion to dust, dust destruction by supernovae, and gas inflow. \citet{key-I2} successfully reproduced the relation between gas-to-dust ratio and gas metallicity at $z\sim0$, and their model predicts a clear evolution of gas-to-dust ratios; at $z\sim1.4$ the gas-to-dust ratio in a galaxy with solar metallicity is expected to be about ten times larger than that at $z\sim0$.

In this study, we aim to derive the gas and dust masses independently to constrain the gas-to-dust ratio in massive star-forming galaxies in the main sequence at $z\sim1.4$. The molecular gas mass is derived from CO($J=1-0$) luminosity and CO-to-H$_2$ conversion factor ($\alpha_\mathrm{CO}$). The CO($J=1-0$) luminosity is derived from CO($J=2-1$) by applying an excitation correction.
We use CO($J=2-1$) emission line to reduce the uncertainty in converting to CO($J=1-0$) luminosity as compared with using higher CO transition lines. Furthermore, we select galaxies with solar metallicity derived from the optical emission line ratio to avoid the uncertainty of $\alpha_\mathrm{CO}$ (see section \ref{sec: sample}). To derive the dust mass, we select galaxies which are detected with {\it Spitzer}/MIPS in 24 $\mu\mathrm{m}$ and {\it Herschel}/SPIRE in 250 $\mu\mathrm{m}$ and 350 $\mu\mathrm{m}$. In Section 2, our sample is described. In Section 3, observations and data reduction are described, and results are presented in Section 4. A discussion and summary are given in Section 5.
Throughout this paper, we adopt the standard $\Lambda$-CDM cosmology with $H_0 = 71\ \mathrm{km\ s^{-1}\ Mpc^{-1}}$, $\Omega_\mathrm{M} = 0.27$, and $\Omega_{\Lambda} = 0.73$.

\section{Sample}
\label{sec: sample}
To reduce the uncertainty of $\alpha_\mathrm{CO}$, we select galaxies with solar metallicity derived based on an optical emission line ratio method (e.g. N2 method: \cite{key-P1}). The value of $\alpha_\mathrm{CO}$ is known to correlate with gas metallicity in local galaxies; the value of $\alpha_\mathrm{CO}$ is larger in galaxies with lower metallicity (e.g., \cite{key-A1}; \cite{key-L1}). A similar relation is found in galaxies at $z = 1 - 2$ (\cite{key-G1}). Although the uncertainty of $\alpha_\mathrm{CO}$ is large in low metallicity galaxies, the value of $\alpha_\mathrm{CO}$ for the solar metallicity is rather solid. According to \citet{key-D1} and \citet{key-G1}, even at $z=1\sim2$, $\alpha_\mathrm{CO}$ is close to the Galactic value ($\alpha_\mathrm{CO} = 4.36\ M_\odot\ (\mathrm{K\ km\ s^{-1}\ pc^2)^{-1}}$, including helium) around solar metallicity.  
We selected the star-forming galaxies with near solar metallicity from the samples of \authorcite{key-Y1} (\yearcite{key-Y1}, \yearcite{key-Y2}) and \citet{key-R1}. \authorcite{key-Y1} (\yearcite{key-Y1}, \yearcite{key-Y2}) and \citet{key-R1} made near infrared spectroscopic observations of star-forming galaxies at $z \sim 1.5$ in the Subaru XMM/Newton Deep Survey (SXDS) field and in the Cosmological Evolution Survey (COSMOS) field, respectively, with Fiber Multi Object Spectrograph (FMOS) on the Subaru telescope.
The gas metallicity is derived from N2 method by using H$\alpha$ and [NII]$\lambda$ 6584 emission lines. 

We further require that the galaxies are detected with {\it Spitzer}/MIPS in 24 $\mu\mathrm{m}$ and {\it Herschel}/SPIRE in 250 $\mu\mathrm{m}$ and 350 $\mu\mathrm{m}$. This allows us to derive the FIR luminosity and the dust mass. For the galaxies in the SXDS field, MIPS data are taken from the DR2 version of Spitzer Public Legacy Survey of the UKIDSS Ultra Deep Survey (SpUDS; Dunlop et al. in preparation). SPIRE images are taken from the DR1 version of the {\it Herschel} Multi-tired Extragalactic Survey (HerMES; \cite{key-O2}). Object detection and photometry are carried out with SExtractor (\cite{key-B2}).  
For the galaxies in the COSMOS field, the photometric data of MIPS and SPIRE were taken from \citet{key-R1}. We made SED from MIR to FIR and derived $L_\mathrm{FIR}(8-1000\ \mu\mathrm{m})$ by fitting model SEDs of star-forming galaxies (\cite{key-C3}).

After these selections, we chose the galaxies which are around the main-sequence of star-forming galaxies (\cite{key-D1}). For the galaxies in the SXDS field, the stellar masses and SFRs are taken from \authorcite{key-Y1} (\yearcite{key-Y1}, \yearcite{key-Y2}); the stellar masses are derived by fitting the spectral energy distribution (SED), and the SFRs are derived from rest-frame UV luminosity density corrected for the dust extinction estimated from the rest-frame UV-slope. For the galaxies in the COSMOS field, the stellar masses and SFRs are taken from \citet{key-W2} after correcting for the difference of initial mass functions (IMFs); the stellar masses are derived by SED fitting and SFRs are derived from far-infrared luminosity and rest-frame UV luminosity.
From these galaxies, we selected three galaxies without contamination by nearby objects in the 250 $\mu$m image and with the largest flux densities in 250 $\mu\mathrm{m}$. The target galaxies selected are summarized in Table \ref{tab: source} and Table \ref{tab: far-IR of source}. 
Although the SFR/(SFR)$_\mathrm{MS}$ of SXDS1\_12778 is small, the stellar mass and SFR of this source are broadly consistent with being within the dispersion of the main sequence. It is noted that if we estimate the SFR from far-infrared luminosity and rest-frame UV luminosity, this source is on the main-sequence (SFR/(SFR)$_\mathrm{MS}\sim1.0$).
Although the metallicity of COSMOS\_9 is only an upper limit (8.68), since the galaxy is massive ($M_\ast \sim 2.5\times 10^{11}\ M_\odot$) the real value of the metallicity is expected to be near solar according to the mass-metallicity relation.

\begin{table*}
  \caption{Sample galaxies.}\label{tab: source}
  \begin{center}
    \begin{tabular}{lccccccc}
      \hline
  Source & R.A. & Decl. & $z_\mathrm{spec}$ & $M_\ast$\footnotemark[$*$] & SFR\footnotemark[$*$] & $\mathrm{\frac{SFR}{(SFR)_{MS}}}$\footnotemark[$\dagger$] & metallcity \\ 
   & (J2000) & (J2000) &  & ($M_\odot$) & ($M_\odot\ \mathrm{yr^{-1}}$) &  & ($12+\log\mathrm{(O/H)}$) \\ 
  \hline
  SXDS1\_12778 & \timeform{2h19m09.45s} & \timeform{-5D09'49.0''} & $1.396$ & $5.6\times10^{11}$ & 90 & 0.10 & $8.66^{+0.09}_{-0.13}$ \\
  SXDS3\_80799 & \timeform{2h17m30.04s} & \timeform{-5D24'31.6''} & $1.429$ & $1.1\times10^{11}$ & 233 & 1.0 & $8.66^{+0.02}_{-0.02}$ \\
  COSMOS\_9 & \timeform{10h00m08.76s} &  \timeform{+2D19'02.3''} & $1.461$ & $2.5\times10^{11}$\footnotemark[$\ddagger$] & 312\footnotemark[$\S$] & 0.68 & $<8.68$ \\
  \hline
 \multicolumn{8}{l}{\parbox{180mm}{\footnotesize
 \par\noindent
 \footnotemark[$*$] We adopted Salpeter IMF (\cite{key-S5}).
 \par\noindent
 \footnotemark[$\dagger$] SFRs normalized by that of the main-sequence galaxies at $z\sim2$ (\cite{key-D1}).
  \par\noindent
 \footnotemark[$\ddagger$] Converted from Chabrier IMF to Salpeter IMF by multiplying 1.8 (\cite{key-E2})
  \par\noindent
 \footnotemark[$\S$] Converted from Kroupa IMF to Salpeter IMF by multiplying 1.6 (\cite{key-F2})
 }}
  \end{tabular}
  \end{center}
\end{table*}

\begin{table}
  \caption{Far-infrared properties of the sample galaxies.}\label{tab: far-IR of source}
  \begin{center}
    \begin{tabular}{lcccc}
      \hline
Source & $S_{24}$ & $S_{250}$ & $S_{350}$ & $L_\mathrm{FIR}$ \\ 
 & (mJy) & (mJy) & (mJy) & ($L_\odot$) \\ 
 \hline
  SXDS1\_12778 & $0.46\pm0.02$ & $45\pm7$ & $59\pm8$ & $3.8\times10^{12}$ \\
  SXDS3\_80799 & $0.59\pm0.02$ & $54\pm6$ & $46\pm5$ & $5.9\times10^{12}$ \\
  COSMOS\_9 & $0.44\pm0.01$ & $26\pm3$ & $25\pm3$ & $4.2\times10^{12}$ \\
  \hline
  \end{tabular}
  \end{center}
\end{table}

\section{Observations and Data Reduction}
We carried out ${}^{12}$CO($J = 2-1$) line observations of the three star-forming galaxies at $z \sim 1.4$. The observed line frequencies shifted from the rest-frame frequency of 230.538 GHz are 93.677 GHz to 96.218 GHz based on the spectroscopic redshifts obtained from the H$\alpha$ observations. 
The CO observations were made on 2013 March 23-26 and May 16-18 with the Nobeyama Radio Observatory 45m telescope. We used the two-beam dual-polarization sideband-separating SIS receiver (TZ receiver: \cite{key-N1}). The beam size (FWHM) was $\sim 17''$ in the frequency range. The backend system was the flexible FX-type spectrometer (e.g., \cite{key-I1}). It had 16 arrays and allowed us to select a bandwidth from several modes between 16 MHz and 2 GHz per array. 
We adopted 2 GHz mode for our observations to cover a wide velocity range of $\sim 6300\ \mathrm{km\ s^{-1}}$. We checked pointing accuracy every 50 minutes by observing SiO maser sources with a 43 GHz HEMT receiver. The pointing accuracy was within $4''$ during the observations. The system noise temperature ($T_\mathrm{sys}$) was typically 160-240 K.

Data reduction was carried out with the NEWSTAR software package. We used only the data which were taken under a wind speed less than 5 m s$^{-1}$. In addition, we flagged scans with poor baselines by inspecting each spectrum by eye. We used three flagging criteria in order to check the robustness of our result.
The image rejection ratios in the adopted frequency range were mostly more than 10 dB, and hence no correction was made. The effective integration time for each galaxy was 2.2-4.5 hours after flagging. 
The observed antenna temperature ($T_\mathrm{A}$) was converted into the main beam brightness temperature ($T_\mathrm{mb}$) using the main beam efficiency of 0.36. Typical rms noise temperature in $T_\mathrm{mb}$ scale was 1.6-2.2 mK after binning to 50 km s$^{-1}$ resolution.

\section{Results}

\subsection{CO (2-1) Spectra}
\label{CO spectra}
\begin{figure*}
  \begin{center}
    \includegraphics[width=17cm]{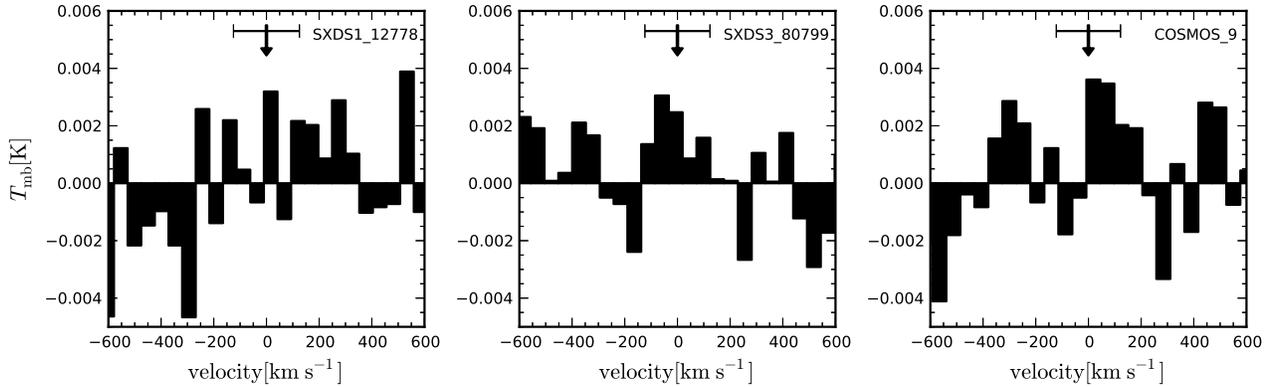}
  \end{center}
  \caption{CO(2-1) spectra taken with the Nobeyama 45m telescope binned with $50\ \mathrm{km\ s^{-1}}$ velocity width. Arrows show the velocity zero points for CO lines expected from spectroscopic redshifts by H$\alpha$ observations. Horizontal bars show the uncertainty of the velocity zero points.} \label{fig:CO spectrum}
\end{figure*}

The spectra obtained are shown in Figure \ref{fig:CO spectrum}.  
The arrow in each panel shows a velocity zero point obtained from the spectroscopic redshift of the H$\alpha$ observations.
The horizontal bar shows the uncertainty of the velocity zero point due to the uncertainty of the redshift. 
In the spectra of SXDS3\_80799 and COSMOS\_9, a weak signal-like feature is seen close to the expected zero point with a velocity width of $\sim 200-250\ \mathrm{km\ s^{-1}}$. The spectrum of SXDS1\_12778 also shows a weak signal-like feature with a $\sim 200\ \mathrm{km\ s^{-1}}$ offset from the arrow. The noise level at $250\mathrm{\ km\ s^{-1}}$ resolution on $T_\mathrm{mb}$ scale, $\sigma_{250}$, is 1.0 mK, 0.7 mK, and 1.0 mK for SXDS1\_12778, SXDS3\_80799, and COSMOS\_9, respectively.
The signal-to-noise ratio (SN) for SXDS1\_12778, SXDS3\_80799, and COSMOS\_9 with $\sim 250\ \mathrm{km\ s^{-1}}$ resolution is 1.8, 2.4, and 2.6, respectively. The signal-like features may be real, but the SN is not good enough to be significant. Hence, we don't regard these features as significant signal and we put upper limits on the CO(2-1) fluxes of the targets.

\subsection{Upper Limit on the Molecular Gas Mass}
\label{Mgas}
The CO line luminosity ($L_\mathrm{CO(1-0)}^{'}$) is given as follows (\cite{key-S2}):
\begin{equation}\label{Lco}
L_\mathrm{CO(1-0)}^{'} = 3.25 \times 10^7\ S_\mathrm{CO(2-1)} \Delta v R_{21}^{-1} \nu_\mathrm{rest(1-0)}^{-2} D_L^2 (1 + z)^{-1},
\end{equation}
where $L_\mathrm{CO(1-0)}^{'}$ is measured in $\mathrm{K\ km\ s^{-1}\ pc^2}$, $S_\mathrm{CO(2-1)}$ is the observed CO$(2-1)$ flux density in Jy, $\Delta v$ is the velocity width in $\mathrm{km\ s^{-1}}$, $R_{21}$ is the CO$(2-1)$/CO$(1-0)$ flux ratio, and $D_L$ is the luminosity distance in Mpc.
We assume a velocity width of $250\ \mathrm{km\ s^{-1}}$, which is hinted from the signal-like features and stacking analysis as describe below (section \ref{stacking}). We smooth the spectrum with a $250\mathrm{\ km\ s^{-1}}$ resolution and take the $2 \sigma_{250}$ upper limit. The value of $R_{21}$ is assumed to be 3, which is a typical value for color-selected star-forming galaxies at $z=1-3$ (e.g., \cite{key-C2}).
The $L_\mathrm{CO(1-0)}^{'}$ in the sample galaxies are given in Table \ref{tab:phot data}. Since the CO($2-1$) signal-like features are seen, the values of $L_\mathrm{CO(1-0)}^{'}$ would not change if the CO lines are really detected. In the diagram of $L_\mathrm{CO(1-0)}^{'} - L_\mathrm{FIR}$ (\cite{key-D1}; \cite{key-G2}), the observed galaxies lie halfway between main-sequence galaxies and submillimeter galaxies (SMGs) or rather close to the SMGs sequence.

Molecular gas mass is derived by 
\begin{equation}\label{Mgas}
M_\mathrm{mol} = \alpha_\mathrm{CO} L_\mathrm{CO(1-0)}^{'}.
\end{equation}
Since the gas metallicities of our targets are close to the solar metallicity, we adopt the Galactic $\alpha_\mathrm{CO}$ ($4.36\ M_\odot\ (\mathrm{K\ km\ s^{-1}\ pc^2)^{-1}}$). The derived upper limits on the molecular gas mass are $(9.7-14)\times10^{10}\ M_\odot$ and these mass fractions are $19-47\ \%$ (Table {\ref{tab:phot data}}). These values are in a similar range as other main-sequence galaxies with similar stellar masses at $z=1-2$ (\cite{key-D1}; \cite{key-T2}).

\subsection{Dust Mass}
The dust mass is derived as
\begin{equation}
M_d = \frac{S_\mathrm{obs} D_L^2}{(1 + z) \kappa_d (\nu_\mathrm{rest}) B (\nu_\mathrm{rest}, T_d)},\label{eq: dust mass}
\end{equation}
where $S_\mathrm{obs}$ is the observed flux density, $\kappa_d (\nu_\mathrm{rest})$ is the dust mass absorption coefficient in the rest-frame frequency, $T_d$ is the dust temperature, and $B (\nu_\mathrm{rest}, T_d)$ is the Planck function. $\kappa_d$ varies with frequency as $\kappa_d \propto \nu^{\beta}$, where $\beta$ is the emissivity. Since the SN is much better in the 250 $\mu$m data than in 350 $\mu$m, we adopt 250 $\mu\mathrm{m}$ flux density for $S_\mathrm{obs}$  which corresponds to $\sim 105\ \mu\mathrm{m}$ in the rest-frame wavelength. We adopt $\kappa_d (125\ \mu \mathrm{m}) = 1.875\ \mathrm{m^2\ kg^{-1}}$ (\cite{key-H1}), and $\beta=1.5$. 
\citet{key-M3} derived the mean dust temperature of star-forming galaxies at $z\sim1.5$ in the $\mathrm{SFR}-M_\ast$ parameter space. 
According to their results in the $\mathrm{SFR}-M_\ast$ diagram, our targets are expected to have dust temperatures of 25-40 K. We adopt $T_d = 35\ \mathrm{K}$. The derived dust masses are $(3.4-6.7)\times10^{8}\ M_\odot$ and are given in Table {\ref{tab:phot data}}.

\begin{table*}
  \caption{Molecular gas and dust properties of the sample galaxies.}\label{tab:phot data}
  \begin{center}
    \begin{tabular}{lccccc}
      \hline
      Source & $L_\mathrm{CO(1-0)}^{'}$ & $M_\mathrm{mol}$\footnotemark[$*$] & $f_\mathrm{mol}$ & $M_d$ & gas-to-dust ratio \\
      & ($\mathrm{K\ km\ s^{-1}\ pc^2}$) & ($M_\odot$) &  & ($M_\odot$) & \\ \hline
      SXDS1\_12778 & $<2.9\times10^{10}$ &  $<1.3 \times 10^{11}$ & $<0.19$ & $5.4 \times 10^{8}$ & $<240$ \\
      SXDS3\_80799 & $<2.2\times10^{10}$ & $<9.7 \times 10^{10}$ & $<0.47$ & $6.7 \times 10^{8}$ & $<150$\\
      COSMOS\_9 & $<3.1\times10^{10}$ & $<1.4 \times 10^{11}$ & $<0.36$ & $3.4 \times 10^{8}$ & $<410$\\
      \hline
      stacking analysis & $(3.0\pm0.8)\times10^{10}$ & $(1.3\pm0.3) \times 10^{11}$ & $0.30^{+0.23}_{-0.19}$ & $(5.2\pm0.4)\times10^{8}$ & $250\pm60$ \\
      \hline
 \multicolumn{6}{l}{\parbox{150mm}{\footnotesize
 \par\noindent
 \footnotemark[$*$] We adopted $\alpha_\mathrm{CO}=4.36\ M_\odot\ (\mathrm{K\ km\ s^{-1}\ pc^2})^{-1}$.
 }}
    \end{tabular}
  \end{center}
\end{table*}

\subsection{Constraints on the Gas-to-Dust Ratio}
\begin{figure}
  \begin{center}
    \includegraphics[width=10cm]{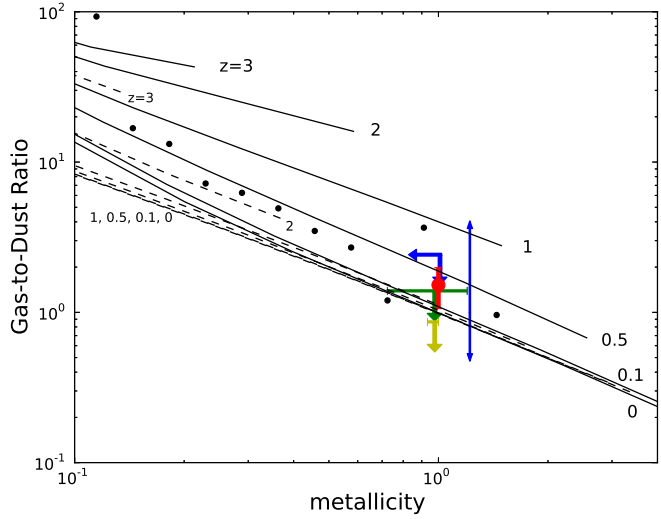}
  \end{center}
  \caption{Gas-to-dust ratios against gas metallicity. Green, yellow, blue, and red symbols refer to SXDS1\_12778, SXDS3\_80799, COSMOS\_9, and the result of stacking analysis, respectively. Arrows show the upper limits and the blue double-headed arrow shows the variation of the limit for COSMOS\_9 with various dust temperatures from 25 K to 40 K. Black circles represent binned results of gas-to-dust ratio in local galaxies (\cite{key-R2}). The solid and dashed lines show expected redshift evolution of gas-to-dust ratio with an accretion time scale of $10^{8}$ yr and $10^{7}$ yr, respectively, taken from the theoretical calculation (\cite{key-I2}). The vertical and horizontal axes are normalized by the gas-to-dust ratio and metallicity of the Galaxy}. \label{fig:GDR}
\end{figure}

We derive the upper limits on the gas-to-dust ratio ($150-410$; Table \ref{tab:phot data}) and plot them against the gas metallicity in Figure \ref{fig:GDR}. The gas-to-dust ratio and metallicity are normalized with the values in the Galaxy, 167 ($=(6\times10^{-3})^{-1}$) and 8.67, respectively. Since we see the signal-like feature in the obtained spectra, the actual gas-to-dust ratio values may be quite close to the limits we can provide based on the upper limits on the CO flux of our galaxies. The blue double-headed arrow in Figure \ref{fig:GDR} shows the variation of the limit of COSMOS\_9 for the dust temperatures from 25 K to 40 K. 

Our limits on the gas-to-dust ratio do not include the contribution from the atomic hydrogen gas. The atomic hydrogen mass is comparable to the molecular gas mass in local massive star-forming galaxies (\cite{key-Y3}). The observations of nearby spiral and dwarf galaxies show that most of the hydrogen is found as molecular hydrogen above an HI critical surface density of $\sim 10\ M_\odot\ \mathrm{pc}^{-2}$ (\cite{key-B1}). 
According to the model of \citet{key-O1}, the atomic hydrogen density is comparable to the molecular hydrogen density in galaxies at $z\sim1.4$. Therefore, our limits of gas-to-dust ratio may be larger by a factor of $\sim 2$, but it does not change the conclusion in this study.

\subsection{Stacking Analysis}\label{stacking}
\begin{figure}
  \begin{center}
    \includegraphics[width=10cm]{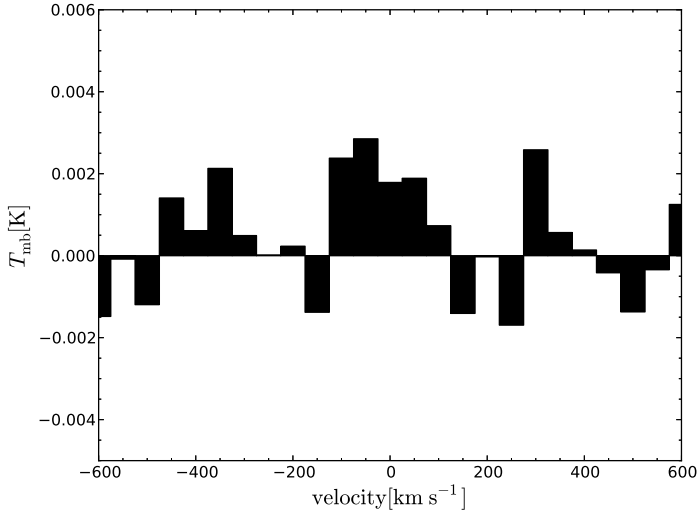}
  \end{center}
  \caption{Stacked CO(2-1) spectrum of three sample galaxies binned with $50\ \mathrm{km\ s^{-1}}$ velocity width. Velocity zero points are shifted to the central positions of the signal-like features.} \label{fig:stacked spectrum}
\end{figure}

Although the central velocities of the signal-like features show offsets from the velocity zero points, the differences are within or comparable to the errors on the velocity. Thus we regard the velocities from the CO observations as most reliable and we stack the three CO spectra by shifting the velocity zero points to the central positions of the signal-like features. The resulting spectrum is shown in Figure \ref{fig:stacked spectrum}. The CO emission is detected significantly; the noise level at $250\ \mathrm{km\ s^{-1}}$ resolution is 0.55 mK and the SN is 3.9. We integrate this CO emission and derive the CO line luminosity of $3.0\times10^{10}\ \mathrm{K\ km\ s^{-1}\ pc^2}$ and the molecular gas mass of $1.3 \times 10^{11}\ M_\odot$ with the same values of $R_{21}$ and $\alpha_\mathrm{CO}$ as described above. 
The average dust mass and $L_\mathrm{FIR}$ of the sample galaxies are $5.2 \times 10^{8}\ M_\odot$ and $4.6\times10^{12}\ L_\odot$, respectively. The data point derived from the stacking analysis in the diagram of $L_\mathrm{CO(1-0)}^{'} - L_\mathrm{FIR}$ lies halfway between main-sequence galaxies and SMGs. The gas-to-dust ratio is 250. We also plot this result in Figure \ref{fig:GDR}. If we stacked the CO spectra at the velocity zero points, the molecular gas mass is $9.9 \times 10^{10}\ M_\odot$ and the gas-to-dust ratio is 190.

\section{Discussion and Summary}
We also plot the expected redshift evolution of the gas-to-dust ratio from theoretical considerations (\cite{key-I2}) in Figure \ref{fig:GDR}. Comparing our results with the theoretical expectation, both our upper limits and the value derived from the stacking analysis lie near the gas-to-dust ratio in galaxies at $z\sim0-0.5$. The range of the limit of COSMOS\_9 caused by the variation of dust temperature shown with the blue double-headed arrow is also higher than the expected gas-to-dust ratio at $z\sim1.4$. Therefore, no significant redshift evolution of the gas-to-dust ratio for massive star-forming galaxies is seen up to $z\sim1.4$ at least for galaxies with solar metallicity.

One possible cause for the non-evolving gas-to-dust ratio is that star-forming galaxies acquired the large dust mass more rapidly. In the model by \citet{key-I2}, one process to increase the dust mass is metal accretion to dust. The time scale of this accretion controls the growth rate of dust mass. Figure \ref{fig:GDR} shows the results for the timescale of $10^{8}\ \mathrm{yr}$, but with the shorter time scale, the dust mass becomes larger rapidly and the gas-to-dust ratio in higher redshift is close to the local value. 
In fact, \citet{key-I2} calculated the evolution of dust-to-gas ratio with shorter time scale of $10^{7}\ \mathrm{yr}$, and the result seems to be consistent with our results (dashed lines in Figure \ref{fig:GDR}). Shattering of dust grains accelerates the growth of the dust mass, and the rapid increase of the dust mass can be realized in an earlier epoch of galaxy evolution (\cite{key-A2}).

Another possible cause is the presence of the gas outflow from a galaxy during the process of galaxy evolution, which is not included in the model by \citet{key-I2}. Gas outflows are observed in many galaxies (e.g., \cite{key-F1}; \cite{key-S3}; \cite{key-C4}). However, dust is also expelled from galaxies by supernovae or radiation pressure, which is suggested from the distribution of dust emissions in the local universe (e.g., \cite{key-K1}; \cite{key-R3}). The outflow presumably does not drive gas outside a galaxy selectively; therefore it may be difficult to reduce the gas-to-dust ratio effectively. 

Since we selected the galaxies which are clearly detected with SPIRE, we should be careful that our sample galaxies may be biased toward sources with large dust mass; the small gas-to-dust ratio may be due to our target selection. 
Observations of molecular gas and dust emission for the star-forming galaxies with lower flux densities in far-infrared are desirable to examine the evolution of gas-to-dust ratio more robustly. 

We carried out $^{12}$CO ($J=2-1$) observations toward three massive star-forming galaxies at $z\sim1.4$ with solar metallicity derived from optical emission lines (N2 method). Our sample galaxies are also detected with {\it Spitzer}/MIPS and {\it Herschel}/SPIRE and are located in the main sequence at $z\sim1.4$. We constrain the gas-to-dust ratios, and find the upper limits at $z\sim1.4$ are near the values for the local galaxies with similar gas metallicity. By stacking the individual CO spectra, we obtain a molecular gas mass and thus gas-to-dust ratio which is comparable to gas-to-dust ratios in local galaxies. Therefore, the results suggest that the gas-to-dust ratio in star-forming galaxies with solar metallicity does not evolve significantly up to $z\sim1.4$. By comparing to a theoretical calculation, this result suggests a rapid increase of the dust mass in an earlier epoch of galaxy evolution.

\bigskip
We would like to thank the referee for useful comments and suggestions.
We acknowledge the members of Nobeyama Radio Observatory for the help during the observations. 
We are grateful to Akio Inoue and Hiroyuki Hirashita for useful discussions. 
We thank Isaac. G. Roseboom for providing the precise redshift information. 
KO is supported by the Grant-in-Aid for Scientific Research (C) (24540230) from Japan Society for the Promotion of Science (JSPS).


\end{document}